\documentclass[pra, twocolumn]{revtex4}
 \usepackage{amssymb} \usepackage{graphicx}

\begin{document}
 \title{Spin raising and lowering operators for Rarita-Schwinger fields}

\author{\"{O}zg\"{u}r A\c{c}{\i}k}
\email{ozacik@science.ankara.edu.tr}
\address{Department of Physics,
Ankara University, Faculty of Sciences, 06100, Tando\u gan-Ankara,
Turkey}
\author{\"Umit Ertem}
 \email{umit.ertem@tedu.edu.tr, umitertemm@gmail.com}
\address{TED University, Ziya G\"{o}kalp Caddesi No:48, 06420, Kolej \c{C}ankaya, Ankara, Turkey}

\begin{abstract}

Spin raising and lowering operators for massless field equations constructed from twistor spinors are considered. Solutions of the spin-$\frac{3}{2}$ massless Rarita-Schwinger equation from source-free Maxwell fields and twistor spinors are constructed. It is shown that this construction requires Ricci-flat backgrounds due to the gauge invariance of the massless Rarita-Schwinger equation. Constraints to construct spin raising and lowering operators for Rarita-Schwinger fields are found. Symmetry operators for Rarita-Schwinger fields via twistor spinors are obtained.

\end{abstract}

\maketitle

\section{Introduction}

In four dimensional conformally flat spacetimes, the solutions of the massless field equations for different spins can be mapped to each other by spin raising and lowering procedures \cite{Penrose Rindler}. A spin raising operator is an operator constructed from a twistor spinor and gives a solution of the spin-$\left(s+\frac{1}{2}\right)$ massless field equation from a solution of the spin-$s$ massless field equation. Similarly, a spin lowering operator maps a solution of the spin-$s$ massless field equation to a solution of the spin-$\left(s-\frac{1}{2}\right)$ massless field equation by using twistor spinors. Twistor spinors are special spinors defined as the solutions of the twistor equation on a spin manifold. They appear in various problems of mathematical physics. Supersymmetry generators of both superconformal field theories in curved backgrounds and conformal supergravity theories correspond to twistor spinors \cite{Klare Tomasiello Zaffaroni, Cassani Martelli, Cassani Klare Martelli Tomasiello Zaffaroni, de Medeiros}. They are also used in the construction of extended conformal superalgebras and are related to the conformal hidden symmetries of a background that are conformal Killing-Yano forms \cite{Ertem1, Ertem2, Ertem3, Acik Ertem}. Twistor spinors contain Killing spinors and parallel spinors as special cases which are supersymmetry generators of supersymmetric field theories and supergravity theories. The classification of manifolds admitting twistor spinors in Riemannian and Lorentzian signatures has been investigated in \cite{Baum Friedrich Grunewald Kath, Baum Leitner}. Especially, they exist on conformally flat manifolds in maximal number.

Starting with a twistor spinor, the spin raising and lowering operators can be constructed for massless spin-0 fields which satisfy the conformally covariant Laplace equation, massless spin-$\frac{1}{2}$ fields which satisfy the massless Dirac equation and massless spin-1 fields which satisfy the source-free Maxwell equations \cite{Benn Charlton Kress, Charlton, Benn Kress}. However, for the case of higher spins, the construction is not straightforward and some constraints may arise in the procedure. Massless spin-$\frac{3}{2}$ fields are solutions of  the massless Rarita-Schwinger equation which determines the motion of gravitino particles in supergravity \cite{Freedman Van Proeyen, Castillo}. Rarita-Schwinger fields appear as sources of torsion and curvature in supergravity field equations. They correspond to spinor-valued 1-forms that are in the kernel of the Rarita-Schwinger operator which can be seen as the generalization of the Dirac operator to spin-$\frac{3}{2}$ fields. Spin raising and lowering procedures can allow us to find the solutions of the massless Rarita-Schwinger equation by using spin-1 source-free Maxwell solutions and twistor spinors.

In this paper, we focus on the construction of spin raising, spin lowering, and symmetry operators for massless Rarita-Schwinger fields. We start by writing Rarita-Schwinger field equations in a modern geometrical language \cite{Benn Tucker, Acik, Acik2}. Spin raising and lowering operators between the massless spin-1 and spin-$\frac{3}{2}$ fields are found by using twistor spinors and the constraints for the construction of them are obtained. Spin raising operators are constructed for middle-form Maxwell fields in all even dimensions besides dimension four and it is found that the twistor spinor used in the construction of the spin raising operator must be in the kernel of the spin-1 Maxwell field. However, since the gauge invariance of the massless Rarita-Schwinger equation requires Ricci-flat backgrounds, it is shown that the spin raising operators automatically solve the massless Rarita-Schwinger equation in those backgrounds. Spin lowering operators are constructed for four dimensional fields and a constraint relating the Rarita-Schwinger field with the curvature characteristics of the background is found. We also construct a symmetry operator for Rarita-Schwinger fields by using spin raising and lowering operators which map a solution of the massless Rarita-Schwinger equation to another solution.

The paper is organized as follows.  We define the spin raising and lowering operators for lower spin massless fields in Sec. 2. In Sec. 3, we construct the spin raising and lowering operators for massless spin-$\frac{3}{2}$ fields. A symmetry operator of massless Rarita-Schwinger fields is proposed in Sec. 4. Section 5 concludes the paper. In an appendix, we give the transformation rules between the languages of Clifford calculus and gamma matrices to write the equalities in the paper in an alternative notation.

\section{Spin Changing Operators for Massless Field Equations}

We consider massless and source-free field equations in curved backgrounds written for particles with different spins. For example, massless spin-0 particles satisfy the following conformally generalized Laplace equation in $n$ dimensions
\begin{equation}
\Delta\varphi-\frac{(n-2)}{4(n-1)}{\cal{R}}\varphi=0
\end{equation}
where $\varphi$ is a function, ${\cal{R}}$ is the scalar curvature of the background spacetime and the Laplace-Beltrami operator $\Delta$ is defined as the square of the Hodge-de Rham operator $\displaystyle{\not}d$, which contains the exterior derivative operator $d$ and coderivative operator $\delta$ as
\begin{eqnarray}
\Delta=\displaystyle{\not}d^2=(d-\delta)^2=-d\delta-\delta d\nonumber.
\end{eqnarray}
The operator given in (1) acting on the scalar field $\varphi$ is called the conformally invariant Yamabe operator. For massless spin-$\frac{1}{2}$ particles, the field equation corresponds to the following massless Dirac equation
\begin{equation}
\displaystyle{\not}D\psi=0
\end{equation}
where $\psi$ is a spinor field and the Dirac operator $\displaystyle{\not}D$ is defined as
\begin{eqnarray}
\displaystyle{\not}D=e^a.\nabla_{X_a}\nonumber
\end{eqnarray}
for the frame basis $\{X_a\}$ and coframe basis $\{e^a\}$ that are related by the duality property $e^a(X_b)=\delta^a_b$. Here $\delta^a_b$ is the Kronecker delta, $\nabla_X$ is the spinor covariant derivative with respect to the vector field $X$ and $.$ denotes the Clifford multiplication. On the other hand, we also consider source-free Maxwell equations for spin-1 particles which are written in terms of the Hodge-de Rham operator $\displaystyle{\not}d=d-\delta$ as follows:
\begin{equation}
\displaystyle{\not}dF=0
\end{equation}
where $F$ is the Maxwell 2-form field \cite{Benn Tucker}.

One can use twistor spinors to obtain solutions of the massless and source-free field equations written in (1), (2) and (3) from the solutions of each other. A twistor spinor $u$ is a solution of the following differential equation in $n$ dimensions:
\begin{equation}
\nabla_X u=\frac{1}{n}\widetilde{X}.\displaystyle{\not}D u
\end{equation}
for any vector field $X$ and its metric dual $\widetilde{X}$. By taking the second covariant derivative of (4), one obtains the following integrability conditions of the twistor equation \cite{Baum Friedrich Grunewald Kath, Benn Kress}:
\begin{eqnarray}
\nabla_{X_a}\displaystyle{\not}D u&=&\frac{n}{2}K_a.u\\
\displaystyle{\not}D^2u&=&-\frac{n}{4(n-1)}{\cal{R}}u\\
C_{ab}.u&=&0
\end{eqnarray}
where the 1-form $K_a$ is defined in terms of the Ricci 1-forms $P_a$ and curvature scalar $\cal{R}$ as
\begin{equation}
K_a=\frac{1}{n-2}\left(\frac{\cal{R}}{2(n-1)}e_a-P_a\right)
\end{equation}
and its components correspond to the Schouten tensor. $C_{ab}$ are conformal 2-forms which are written in terms of the curvature characteristics and defined for $n>2$ as
\begin{eqnarray}
C_{ab}&=&R_{ab}-\frac{1}{n-2}\left(P_a\wedge e_b-P_b\wedge e_a\right)\nonumber\\
&&+\frac{1}{(n-1)(n-2)}{\cal{R}}e_a\wedge e_b
\end{eqnarray}
where $R_{ab}$ are curvature 2-forms and $\wedge$ denotes the wedge product. The components of the 2-form $C_{ab}$ correspond to the conformal (Weyl) tensor. The second integrability condition given in (6) is the standard Weitzenbock identity. One can see from the third integrability condition in (7) that all conformally flat manifolds can have solutions of the twistor equation (4).

In four dimensions, a solution $\psi$ of the massless Dirac equation given in (2) can be constructed from a solution $\varphi$ of the massless spin-0 particle equation in (1) and a twistor spinor $u$ that satisfies (4) \cite{Benn Charlton Kress, Benn Kress}. The so-called "spin raising" operator from spin-0 to spin-$\frac{1}{2}$ is given as follows:
\begin{equation}
\psi=d\varphi.u+\frac{1}{2}\varphi.\displaystyle{\not}Du.
\end{equation}
It can easily be seen that by using the defining equations of $\varphi$ and $u$ given by (1) and (4) and the integrability conditions (5)-(7), one can find the result $\displaystyle{\not}D\psi=0$. A reverse procedure of obtaining a solution $\varphi$ of massless spin-0 equation from a solution $\psi$ of the massless Dirac equation and a twistor spinor $u$ can also be constructed \cite{Charlton, Benn Kress}. To do this, we consider the spin-invariant inner product $(\,,\,)$ defined on spinor fields \cite{Benn Tucker}. For any two spinor fields $u$ and $v$, it has the property $(u,v)=-(v,u)$ and for any differential form $\alpha$, we have
\begin{eqnarray}
(u,\alpha.v)=(\alpha^{\xi}.u,v)\nonumber
\end{eqnarray}
where $\alpha^{\xi}=(-1)^{\lfloor p/2\rfloor}\alpha$ with $\lfloor\rfloor$ denoting the floor function that takes the integer part of its argument. By using this inner product, the so-called "spin lowering" operator from spin-$\frac{1}{2}$ to spin-0 is given in the following form:
\begin{equation}
\varphi=(u,\psi).
\end{equation}
One can check by using the defining equations (2) and (4) and the integrability conditions (5)-(7) that this function $\varphi$ satisfies the massless spin-0 field equation given in (1). By combining the spin lowering and spin raising operations, one can also construct symmetry operators for massless Dirac fields. A symmetry operator takes a solution of an equation and gives another solution. So, by considering two twistor spinors $u_1$ and $u_2$ and a solution $\psi$ of the massless Dirac equation, the following symmetry operator for massless Dirac fields can be written from the spin lowering and spin raising operations
\begin{equation}
L_{u_1u_2}\psi=d(u_1,\psi).u_2+\frac{1}{2}(u_1,\psi).\displaystyle{\not}Du_2.
\end{equation}
Since the antisymmetric generalizations of conformal Killing vector fields that are called conformal Killing-Yano forms can be constructed from two twistor spinors \cite{Acik Ertem}, it can be shown that the symmetry operator (12) is equivalent to the symmetry operators of massless Dirac equation written in terms of conformal Killing-Yano forms \cite{Benn Charlton Kress, Benn Charlton}.

Spin raising and spin lowering operators can also be defined for obtaining solutions of spin-1 source-free Maxwell equations from the solutions of the massless Dirac equation and vice versa. To do this, we consider the dual spinor $\overline{u}$ of a spinor $u$. The dual spinor $\overline{u}$ is defined by the action of it on a spinor $v$ in terms of the spinor inner product as follows:
\begin{eqnarray}
\overline{u}(v)=(u,v).\nonumber
\end{eqnarray}
We can consider tensor products of spinors and dual spinors which correspond to the linear transformations on spinors
\begin{eqnarray}
(u\otimes\overline{v})w=u(v,w)\nonumber
\end{eqnarray}
and 
\begin{eqnarray}
(u\otimes\overline{v})\overline{w}=(w,u)\overline{v}\nonumber
\end{eqnarray}
for any spinor $w$ \cite{Charlton}. Since the inner products $(v,w)$ and $(w,v)$ give scalar quantities and $u$ and $\overline{v}$ are a spinor and a dual spinor, respectively, the quantities $u(v,w)$ and $(w,u)\overline{v}$ correspond to a spinor and a dual spinor respectively. This means that tensor products of spinors and dual spinors are elements of the Clifford algebra and they can be Clifford multiplied by inhomogeneous differential forms. For any differential form $\alpha$, we have the following relations for the tensor product of spinors and dual spinors:
\begin{eqnarray}
\alpha.(u\otimes\overline{v})&=&\alpha.u\otimes\overline{v}\nonumber\\
(u\otimes\overline{v}).\alpha&=&u\otimes\overline{\alpha^{\xi}.v}\nonumber\\
(u\otimes\overline{v})^{\xi}&=&-v\otimes\overline{u}\nonumber.
\end{eqnarray}
So, from a solution $\psi$ of the massless Dirac equation and a twistor spinor $u$, one can write the following spin raising operator from spin-$\frac{1}{2}$ to spin-1 as follows \cite{Benn Charlton Kress, Charlton}:
\begin{equation}
F=e^a.u\otimes\overline{\nabla_{X_a}\psi}+\frac{1}{2}\displaystyle{\not}Du\otimes\overline{\psi}+\psi\otimes\overline{\displaystyle{\not}Du}.
\end{equation}
It can be seen that from the properties of the tensor product given above, the differential form $F$ in (13) is a 2-form in four dimensions and from the defining equations of (2) and (4) and the integrability conditions (5)-(7), it satisfies the source-free Maxwell equation (3) \cite{Benn Charlton Kress, Charlton}. Similarly, we can also construct a spin lowering operator from a solution $F$ of the source-free Maxwell equations and a twistor spinor $u$ to obtain a solution of the massless Dirac equation. The spin lowering operator from spin-1 to spin-$\frac{1}{2}$ is given as
\begin{equation}
\psi=F.u
\end{equation}
It can be checked that $\psi$ in (14) is a harmonic spinor, namely, a solution of the massless Dirac equation. From the spin lowering and spin raising operations between the solutions of spin-1 and spin-$\frac{1}{2}$ field equations, the symmetry operators of source-free Maxwell equations can be constructed in the following form:
\begin{equation}
L_{u_1u_2}F=e^a.u_2\otimes\overline{\nabla_{X_a}(F.u_1)}+\frac{1}{2}\displaystyle{\not}Du_2\otimes\overline{F.u_1}+F.u_1\otimes\overline{\displaystyle{\not}Du_2}
\end{equation}
where $u_1$ and $u_2$ are two twistor spinors and $F$ is a source-free Maxwell solution.

\section{Spin Raising and Spin Lowering for Rarita-Schwinger Fields}

In this section, we construct spin raising and spin lowering operators for spin-$\frac{3}{2}$ fields that satisfy the massless Rarita-Schwinger field equations. These operators will be used to make connections between the solutions of the source-free Maxwell equations and the massless Rarita-Schwinger equations via twistor spinors in all even dimensions besides the special case of dimension four.

Let us first consider spinor-valued $p$-forms which are constructed out of tensor products of spinor fields and differential $p$-forms and describe spin-$\left(p+\frac{1}{2}\right)$ particles. For a spinor with indices $\psi_I$ and a $p$-form $e^I$, the spinor-valued $p$-form is defined by
\begin{eqnarray}
\Psi_p=\psi_I\otimes e^I\nonumber
\end{eqnarray}
where $I$ is a multi-index. Note that, the tensor product of spinor fields and $p$-forms defined above is different from the tensor product of spinors and dual spinors defined in Sec. 2 although we denote them with the same symbol. In a similar way, we can also define a Clifford form-valued $p$-form $N_p=n_A\otimes e^A$ as a tensor product of a differential form $n_A$ on the Clifford bundle and a differential $p$-form $e^A$ on the exterior bundle. The action of a Clifford form-valued $p$-form $N_p$ to the spinor-valued $q$-form $\Psi_q$ is defined by
\begin{eqnarray}
N_p.\Psi_q=n_A.\psi_I\otimes e^A\wedge e^I
\end{eqnarray}
in terms of Clifford and wedge products, the result of which is a spinor-valued $(p+q)$-form. As a special case, we consider spinor-valued 1-forms representing spin-$\frac{3}{2}$ particles. In that case, we have a spinor field $\psi_a$ and the coframe basis $e^a$ of 1-forms to construct the spin-$\frac{3}{2}$ quantity
\begin{eqnarray}
\Psi=\psi_a\otimes e^a\nonumber.
\end{eqnarray}
The action of a Clifford form $\alpha$ on $\Psi$ is defined by
\begin{eqnarray}
\alpha.\Psi=\alpha.\psi_a\otimes e^a.
\end{eqnarray}
Moreover, the inner product of a spinor-valued 1-form $\Psi$ and a spinor $u$ in terms of the spinor inner product $(\,,\,)$ is given as follows
\begin{equation}
(u,\Psi)=(u,\psi_a)e^a
\end{equation}
which takes a spinor and a spinor-valued 1-form and gives a 1-form.

The Levi-Civita connection defined on differential forms and spinor fields can also be induced on spinor-valued 1-forms with the following property:
\begin{equation}
\nabla_X\Psi=\nabla_X\psi_a\otimes e^a+\psi_a\otimes\nabla_Xe^a
\end{equation}
for any vector field $X$. In a similar way to the definition of the Dirac operator on spinor fields, we can also define a Rarita-Schwinger operator
\begin{eqnarray}
{\mathbb{\displaystyle{\not}D}}=e^a.\nabla_{X_a}
\end{eqnarray}
which acts on spinor-valued 1-forms. From the definitions given above, the massless Rarita-Schwinger field equations of spin-$\frac{3}{2}$ fields in supergravity can be written for a spin-$\frac{3}{2}$ field $\Psi=\psi_a\otimes e^a$ as follows
\begin{eqnarray}
{\mathbb{\displaystyle{\not}D}}\Psi&=&0\\
e^a.\psi_a&=&0.
\end{eqnarray}
Equation (21) can be seen as a generalization of the massless Dirac equation to spin-$\frac{3}{2}$ fields and Eq.(22) is the tracelessness condition. Moreover, these equations imply a Lorentz-type condition $\nabla_{X^a}\psi_a=0$. This can be seen as follows. By taking the covariant derivative of the Rarita-Schwinger field $\Psi=\psi_a\otimes e^a$
\begin{eqnarray}
\nabla_{X_b}\Psi=\nabla_{X_b}\psi_a\otimes e^a+\psi_a\otimes\nabla_{X_b}e^a\nonumber.
\end{eqnarray}
We will use $\nabla_{X_b}e^a=0$ for normal coordinates in the following calculations. By Clifford multiplying with $e^b$ from the left and using (21), we find
\begin{eqnarray}
\displaystyle{\not}D\psi_a\otimes e^a=0\nonumber.
\end{eqnarray}
Then, we have $\displaystyle{\not}D\psi_a=0$ for every $a$. Again, by Clifford multiplying with $e^a$ from the left and using the Clifford algebra identity $e^a.e^b+e^b.e^a=2g^{ab}$ for the components of the (dual) metric $g^{ab}$, we obtain
\begin{eqnarray}
e^a.e^b.\nabla_{X_b}\psi_a&=&2\nabla_{X^a}\psi_a-e^b.e^a.\nabla_{X_b}\psi_a\nonumber\\
&=&2\nabla_{X^a}\psi_a-e^b.\nabla_{X_b}(e^a.\psi_a)=0\nonumber
\end{eqnarray}
where we have used $\nabla_{X_b}e^a=0$ in normal coordinates and from (22) we obtain the following Lorentz-type condition
\begin{equation}
\nabla_{X^a}\psi_a=0.
\end{equation}

\subsection{Spin raising}

Let us consider a twistor spinor $u$ which satisfies Eq. (4) and a middle-form Maxwell field $F$ which satisfies Eq. (3). For even dimensions $n=2p$, the $p$-form Maxwell field $F$ is the generalization of the 2-form Maxwell field in four dimensions to all even dimensions. We propose the following spinor-valued 1-form that is the spin-$\frac{3}{2}$ field as the spin raising operator from the spin-1 Maxwell field $F$ to the spin-$\frac{3}{2}$ Rarita-Schwinger field $\Psi=\psi_a\otimes e^a$ via the twistor spinor $u$
\begin{equation}
\Psi=\left(\nabla_{X_a}F.u-\frac{1}{n}F.e_a.\displaystyle{\not}Du\right)\otimes e^a.
\end{equation}
 Here it is clear that $\psi_a=\left(\nabla_{X_a}F.u-\frac{1}{n}F.e_a.\displaystyle{\not}Du\right)$. Now, we have to prove that the spin-$\frac{3}{2}$ field defined in (24) satisfies both of the Rarita-Schwinger field equations in (21) and (22).

Using this $\psi_a$ in (22) we obtain
\begin{eqnarray}
e^a.\psi_a=e^a.\nabla_{X_a}F.u-\frac{1}{n}e^a.F.e_a.\displaystyle{\not}Du\nonumber.
\end{eqnarray}
For a $p$-form $\alpha$, we have the relation $e^a.\alpha.e_a=(-1)^p(n-2p)\alpha$. From this identity and the definition of the Hodge-de Rham operator $\displaystyle{\not}d$ acting on any differential form $\alpha$ as $\displaystyle{\not}d\alpha=e^a.\nabla_{X_a}\alpha$, we find
\begin{eqnarray}
e^a.\psi_a=\displaystyle{\not}dF.u-(-1)^p\frac{n-2p}{n}F.\displaystyle{\not}Du.
\end{eqnarray}
Since $F$ satisfies $\displaystyle{\not}dF=0$ and we have $n=2p$, one concludes
\begin{eqnarray}
e^a.\psi_a=0\nonumber.
\end{eqnarray}
As a consequence, $\Psi$ defined in (24) satisfies the tracelessness condition of a Rarita-Schwinger field.

To see if Eq.(21) is satisfied by $\Psi$ in (24), we apply the Rarita-Schwinger operator defined in (20) to $\Psi$
\begin{eqnarray}
\displaystyle{\not}\mathbb{D}\Psi&=&e^b.\nabla_{X_b}\left[\left(\nabla_{X_a}F.u-\frac{1}{n}F.e_a.\displaystyle{\not}Du\right)\otimes e^a\right]\nonumber\\
&=&e^b.\bigg(\nabla_{X_b}\nabla_{X_a}F.u+\nabla_{X_a}F.\nabla_{X_b}u\\
&&-\frac{1}{n}\nabla_{X_b}F.e_a.\displaystyle{\not}Du-\frac{1}{n}F.e_a.\nabla_{X_b}\displaystyle{\not}Du\bigg)\otimes e^a\nonumber
\end{eqnarray}
where we have used (19) and $\nabla_{X_b}e^a=0$ in normal coordinates. From the property (17) and the definition of $\displaystyle{\not}d$, we obtain
\begin{eqnarray}
\displaystyle{\not}\mathbb{D}\Psi&=&\bigg(\displaystyle{\not}d\nabla_{X_a}F.u+e^b.\nabla_{X_a}F.\nabla_{X_b}u-\frac{1}{n}\displaystyle{\not}dF.e_a.\displaystyle{\not}Du\nonumber\\
&&-\frac{1}{n}e^b.F.e_a.\nabla_{X_b}\displaystyle{\not}Du\bigg)\otimes e^a.
\end{eqnarray}
We know that $F$ satisfies (3) and $u$ is a twistor spinor which satisfies (4) and (5). Then, we have
\begin{eqnarray}
\displaystyle{\not}\mathbb{D}\Psi&=&\bigg(\displaystyle{\not}d\nabla_{X_a}F.u+\frac{1}{n}e^b.\nabla_{X_a}F.e_b.\displaystyle{\not}Du\nonumber\\
&&-\frac{1}{2}e^b.F.e_a.K_b.u\bigg)\otimes e^a\nonumber\\
&=&\left(\displaystyle{\not}d\nabla_{X_a}F.u-\frac{1}{2}e^b.F.e_a.K_b.u\right)\otimes e^a
\end{eqnarray}
where we have used $e^b.\nabla_{X_a}F.e_b=(-1)^p(n-2p)\nabla_{X_a}F=0$ since $\nabla_{X_a}F$ is a $\frac{n}{2}$-form. By direct computation, one can see that the action of the commutator of the Hodge-de Rham operator and the covariant derivative on a differential $p$-form $\alpha$ can be written in terms of the curvature operator as
\begin{eqnarray}
[\displaystyle{\not}d,\nabla_{X_a}]\alpha=-e^b.R(X_a,X_b)\alpha.
\end{eqnarray}
By using this identity, (28) transforms into the following form
\begin{eqnarray}
\displaystyle{\not}\mathbb{D}\Psi&=&\bigg(-e^b.R(X_a,X_b)F.u+\nabla_{X_a}\displaystyle{\not}dF.u\nonumber\\
&&-\frac{1}{2}e^b.F.e_a.K_b.u\bigg)\otimes e^a\\
&=&\left(-e^b.R(X_a,X_b)F.u-\frac{1}{2}e^b.F.e_a.K_b.u\right)\otimes e^a\nonumber\\
&=&\left[-e^b.\left(R(X_a,X_b)F+\frac{1}{2}F.e_a.K_b\right).u\right]\otimes e^a\nonumber
\end{eqnarray}
where we have used (3) in the second line.

The action of the curvature operator on a generally inhomogeneous differential form $\alpha$ which is a section of the Clifford bundle can be written in terms of the curvature 2-forms and Clifford bracket as \cite{Benn Tucker}
\begin{equation}
R(X_a,X_b)\alpha=\frac{1}{2}[R_{ab},\alpha]_{Cl}
\end{equation}
where $[\,,\,]_{Cl}$ is defined as $[R_{ab},\alpha]_{Cl}=R_{ab}.\alpha-\alpha.R_{ab}$. Moreover, from the definition of conformal 2-forms $C_{ab}$ in (9), we can write the curvature 2-forms as follows:
\begin{eqnarray}
R_{ab}&=&C_{ab}+\frac{1}{n-2}\left(P_a\wedge e_b-P_b\wedge e_a\right)\nonumber\\
&&-\frac{1}{(n-1)(n-2)}{\cal{R}}e_a\wedge e_b\nonumber\\
&=&C_{ab}+K_b\wedge e_a-K_a\wedge e_b\nonumber\\
&=&C_{ab}+e_b.K_a-e_a.K_b
\end{eqnarray}
where we have used the definition of $K_a$ in (8) and the expansion of the Clifford product in terms of the wedge product and contraction $i_X$ with respect to a vector field $X$ as $e_a.K_b=e_a\wedge K_b+i_{X_a}K_b$ with the property $i_{X_a}K_b=i_{X_b}K_a$ for zero torsion. So, by substituting (31) and (32) in (30), we find
\begin{eqnarray}
\displaystyle{\not}\mathbb{D}\Psi&=&\left(-\frac{1}{2}e^b.[R_{ab},F]_{Cl}.u-\frac{1}{2}e^b.F.e_a.K_b.u\right)\otimes e^a\nonumber\\
&=&\bigg(-\frac{1}{2}e^b.R_{ab}.F.u+\frac{1}{2}e^b.F.R_{ab}.u\nonumber\\
&&-\frac{1}{2}e^b.F.e_a.K_b.u\bigg)\otimes e^a\nonumber\\
&=&\bigg(\frac{1}{2}P_a.F.u+\frac{1}{2}e^b.F.C_{ab}.u+\frac{1}{2}e^b.F.e_b.K_a.u\nonumber\\
&&-e^b.F.e_a.K_b.u\bigg)\otimes e^a\nonumber\\
&=&\left(\frac{1}{2}P_a.F.u-e^b.F.e_a.K_b.u\right)\otimes e^a
\end{eqnarray}
where we have used the identity $e^b.R_{ab}=-P_a$ for zero torsion, $e^b.F.e_b=0$ and the integrability condition (7). Then, for $\Psi$ to satisfy (21), we obtain the condition
\begin{equation}
P_a.F.u=2e^b.F.e_a.K_b.u
\end{equation}
or from the definition (8) of $K_a$, it can also be written as
\begin{equation}
P_a.F.u=\frac{2{\cal{R}}}{(n-1)(n-2)}e_a.F.u-\frac{2}{n-2}e^b.F.e_a.P_b.u.
\end{equation}
By Clifford multiplying (35) with $e^a$ from the left and using the identities $e^a.P_a={\cal{R}}$, $e^a.e_a=n$, $e^a.e^b=-e^b.e^a+2g^{ab}$ and the property $e^a.F.e_a=0$, one obtains the following equality:
\begin{equation}
\left(\frac{n(n-1)-2}{(n-1)(n-2)}\right){\cal{R}}F.u=0.
\end{equation}
So, the condition (35) for $\Psi$ to be a massless Rarita-Schwinger field transforms into the following condition
\begin{equation}
F.u=0.
\end{equation}
This resembles a condition on Killing-Yano forms that can be used in the construction of symmetry operators of a massive Dirac equation with an electromagnetic minimal coupling term \cite{Acik Ertem Onder Vercin}. Those symmetry operators are constructed from the Killing-Yano forms $\omega$ that satisfy the condition
\[
[F,\omega]_{Cl}=0
\]
where $[\,,\,]_{Cl}$ is the Clifford bracket. Since a Killing spinor $u$, which is a special twistor spinor that satisfies the massive Dirac equation at the same time, can be used in the construction of the Killing-Yano form $\omega$ as $\omega=u\otimes\overline{u}$ \cite{Acik Ertem}, the above condition on $\omega$ reduces to the condition on $u$ written in (37).

On the other hand, the gauge invariance of the massless Rarita-Schwinger equation in a curved background requires the background to be Ricci flat, that is $P_a=0$. This can be seen as follows. The Rarita-Schwinger equation given in (21) and (22) can be written in a more compact form. Let us consider a Clifford-valued 1-form $e=e^a\otimes e_a$ and define the action of the covariant exterior derivative $D$ on $\Psi=\psi_a\otimes e^a$ as
\begin{eqnarray}
D\Psi&=&e^a\wedge\nabla_{X_a}(\psi_b\otimes e^b)\nonumber\\
&=&\nabla_{X_a}\psi_b\otimes e^{ab}\nonumber
\end{eqnarray}
where we have used $\nabla_{X_a}e^b=0$ in normal coordinates and $e^{ab}=e^a\wedge e^b$. The action of the Hodge star $*$ on $D\Psi$ is given by $*D\Psi=\nabla_{X_a}\psi_b\otimes *e^{ab}$. From the action of Clifford-valued forms on spinor-valued forms defined in (16), we can write
\begin{eqnarray}
e.*D\Psi&=&e^c.\nabla_{X_a}\psi_b\otimes e_c\wedge*e^{ab}\nonumber\\
&=&(e^a.\nabla_{X_b}\psi_a-\displaystyle{\not}D\psi_b)\otimes*e^b\nonumber\\
&=&\nabla_{X_b}(e^a.\psi_a)\otimes*e^b-\displaystyle{\not}D\psi_b\otimes*e^b\nonumber\\
&=&\nabla_{X_b}(e^a.\psi_a)\otimes*e^b-*\mathbb{\displaystyle{\not}D}\Psi\nonumber
\end{eqnarray}
where we have used the identity $e^c\wedge*e^{ab}=g^{ca}*e^b-g^{cb}*e^a$ in the second line and the action of the Hodge star in the last line. So, from (21) and (22), the Rarita-Schwinger equation is equivalent to
\[
e.*D\Psi=0.
\]
This equation has to be gauge invariant under the transformation $\Psi\rightarrow\Psi+D\phi$ for a spinor $\phi$. We can write the spinor $\phi$ as $\phi=\phi\otimes 1$ and
\begin{eqnarray}
D\phi&=&e^a\wedge\nabla_{X_a}(\phi\otimes 1)\nonumber\\
&=&\nabla_{X_a}\phi\otimes e^a.\nonumber
\end{eqnarray}
By applying $D$ once more, we have
\begin{eqnarray}
D^2\phi&=&e^b\wedge\nabla_{X_b}(\nabla_{X_a}\phi\otimes e^a)\nonumber\\
&=&\nabla_{X_b}\nabla_{X_a}\phi\otimes e^{ba}\nonumber\\
&=&\frac{1}{2}(\nabla_{X_b}\nabla_{X_a}-\nabla_{X_a}\nabla_{X_b})\phi\otimes e^{ba}\nonumber\\
&=&\frac{1}{2}R(X_b,X_a)\phi\otimes e^{ba}\nonumber
\end{eqnarray}
where we have used the antisymmetry of the indices in the third line and the definition of the curvature operator in the fourth line. Then, if we choose $\Psi$ as a pure gauge term $\Psi=D\phi$, we obtain
\begin{eqnarray}
e.*D^2\phi&=&\frac{1}{2}e_c.R(X_b,X_a)\phi\otimes e^c\wedge*e^{ba}\nonumber\\
&=&-e^b.R(X_b,X_a)\phi\otimes*e^a\nonumber\\
&=&-\frac{1}{2}e^b.R_{ba}.\phi\otimes*e^a\nonumber\\
&=&-\frac{1}{2}P_a.\phi\otimes*e^a\nonumber
\end{eqnarray}
where we have used the identity $e^c\wedge*e^{ab}=g^{ca}*e^b-g^{cb}*e^a$ in the second line, $R(X_b,X_a)\phi=\frac{1}{2}R_{ba}.\phi$ in the third line and $e^b.R_{ba}=P_a$ in the last line. The gauge invariance implies the vanishing of the pure gauge term, namely $e.*D^2\phi=0$. So, to obtain a gauge invariant massless Rarita-Schwinger equation we must have a Ricci-flat background, $P_a=0$. In that case, the right-hand side of (33) automatically vanishes and we obtain a Rarita-Schwinger field $\Psi$ from a source-free Maxwell field $F$ and a twistor spinor $u$ as constructed in (24).

\subsection{Spin lowering}

By using a twistor spinor $u$, we can also construct a spin lowering procedure to obtain a spin-1 Maxwell field $F$ from a spin-$\frac{3}{2}$ Rarita-Schwinger field $\Psi=\psi_a\otimes e^a$ in four dimensions. From the inner product definition (18) for spinor-valued 1-forms, let us consider the following 1-form $A$ constructed out of a twistor spinor $u$ and a Rarita-Schwinger field $\Psi=\psi_a\otimes e^a$ satisfying (21) and (22):
\begin{equation}
A=(u,\Psi)=(u,\psi_a)e^a.
\end{equation}
We consider the 1-form $A$ as the potential 1-form of the 2-form Maxwell field $F=dA$. So, we have
\begin{eqnarray}
F&=&d(u,\Psi)\nonumber\\
&=&e^b\wedge\nabla_{X_b}\left[(u,\psi_a)e^a\right]\nonumber\\
&=&\left[(\nabla_{X_b}u,\psi_a)+(u,\nabla_{X_b}\psi_a)\right]e^b\wedge e^a\nonumber\\
&=&\left[\frac{1}{n}(e_b.\displaystyle{\not}Du,\psi_a)+(u,\nabla_{X_b}\psi_a)\right]e^b\wedge e^a
\end{eqnarray}
where we have used normal coordinates, the definition $d=e^b\wedge\nabla_{X_b}$ and the twistor equation (4). By definition, $F$ is an exact form and it automatically satisfies $dF=0$. So the action of Hodge-de Rham operator $\displaystyle{\not}d=d-\delta$ on $F$ gives
\begin{eqnarray}
\displaystyle{\not}dF&=&dF-\delta F\nonumber\\
&=&i_{X^c}\nabla_{X_c}F\nonumber
\end{eqnarray}
from the definition $\delta=-i_{X^c}\nabla_{X_c}$. By taking the covariant derivative of $F$, we find from (39)
\begin{eqnarray}
\nabla_{X_c}F&=&\left[\frac{1}{n}\nabla_{X_c}(e_b.\displaystyle{\not}Du,\psi_a)+\nabla_{X_c}(u,\nabla_{X_b}\psi_a)\right]e^b\wedge e^a\nonumber\\
&=&\bigg[\frac{1}{2n}\nabla_{X_c}(\displaystyle{\not}Du,e_b.\psi_a-e_a.\psi_b)\nonumber\\
&&+\frac{1}{2}\nabla_{X_c}(u,\nabla_{X_b}\psi_a-\nabla_{X_a}\psi_b)\bigg]e^b\wedge e^a
\end{eqnarray}
where we have used the identity $(e_b.\displaystyle{\not}Du,\psi_a)=(\displaystyle{\not}Du,e_b.\psi_a)$ and antisymmetrized the corresponding indices. So, we have
\begin{eqnarray}
\nabla_{X_c}F&=&\bigg[\frac{1}{2n}(\nabla_{X_c}\displaystyle{\not}Du,e_b.\psi_a-e_a.\psi_b)\nonumber\\
&&+\frac{1}{2n}(\displaystyle{\not}Du,e_b.\nabla_{X_c}\psi_a-e_a.\nabla_{X_c}\psi_b)\nonumber\\
&&+\frac{1}{2}(\nabla_{X_c}u,\nabla_{X_b}\psi_a-\nabla_{X_a}\psi_b)\\
&&+\frac{1}{2}(u,\nabla_{X_c}\nabla_{X_b}\psi_a-\nabla_{X_c}\nabla_{X_a}\psi_b)\bigg]e^b\wedge e^a\nonumber.
\end{eqnarray}
From the identity $i_{X^c}(e^b\wedge e^a)=g^{cb}e^a-g^{ca}e^b$, we obtain the action of $\displaystyle{\not}d$ on $F$ as
\begin{eqnarray}
\displaystyle{\not}dF&=&i_{X^c}\nabla_{X_c}F\nonumber\\
&=&\bigg[\frac{1}{2n}(\nabla_{X_c}\displaystyle{\not}Du,e^c.\psi_a-e_a.\psi^c)\nonumber\\
&&+\frac{1}{2n}(\displaystyle{\not}Du,e^c.\nabla_{X_c}\psi_a-e_a.\nabla_{X_c}\psi^c)\nonumber\\
&&+\frac{1}{2}(\nabla_{X_c}u,\nabla_{X^c}\psi_a-\nabla_{X_a}\psi^c)\nonumber\\
&&+\frac{1}{2}(u,\nabla_{X_c}\nabla_{X^c}\psi_a-\nabla_{X_c}\nabla_{X_a}\psi^c)\nonumber\\
&&-\frac{1}{2n}(\nabla_{X_c}\displaystyle{\not}Du,e_a.\psi^c-e^c.\psi_a)\nonumber\\
&&-\frac{1}{2n}(\displaystyle{\not}Du,e_a.\nabla_{X_c}\psi^c-e^c.\nabla_{X_c}\psi_a)\nonumber\\
&&-\frac{1}{2}(\nabla_{X_c}u,\nabla_{X_a}\psi^c-\nabla_{X^c}\psi_a)\nonumber\\
&&-\frac{1}{2}(u,\nabla_{X_c}\nabla_{X_a}\psi^c-\nabla_{X_c}\nabla_{X^c}\psi_a)\bigg]e^a.
\end{eqnarray}
Since $\Psi$ is a Rarita-Schwinger field, we have $e^c.\nabla_{X_c}\psi_a=\displaystyle{\not}D\psi_a=0$ and $\nabla_{X_a}\psi^a=0$. From the twistor equation (4), we can write
\begin{eqnarray}
(\nabla_{X_c}u,\nabla_{X^c}\psi_a-\nabla_{X_a}\psi^c)&=&\frac{1}{n}(e_c.\displaystyle{\not}Du,\nabla_{X^c}\psi_a-\nabla_{X_a}\psi^c)\nonumber\\
&=&\frac{1}{n}(\displaystyle{\not}Du,\displaystyle{\not}D\psi_a-e_c.\nabla_{X_a}\psi^c)\nonumber\\
&=&0\nonumber
\end{eqnarray}
where we have used $e_c.\nabla_{X_a}\psi^c=\nabla_{X_a}(e_c.\psi^c)=0$ from $e_c.\psi^c=0$. Then, (42) transforms into
\begin{eqnarray}
\displaystyle{\not}dF&=&\bigg[\frac{1}{n}(\nabla_{X_c}\displaystyle{\not}Du,e^c.\psi_a)-\frac{1}{n}(\nabla_{X_c}\displaystyle{\not}Du,e_a.\psi^c)\nonumber\\
&&+(u,\nabla_{X_c}\nabla_{X^c}\psi_a)-(u,\nabla_{X_c}\nabla_{X_a}\psi^c)\bigg]e^a.
\end{eqnarray}
By defining the spinor Laplacian $\nabla^2=\nabla_{X_c}\nabla_{X^c}$ and using the Schr\"{o}dinger-Lichnerowicz-Weitzenb\"{o}ck formula for spinor fields which is
\begin{equation}
\displaystyle{\not}D^2=\nabla^2-\frac{1}{4}{\cal{R}}
\end{equation}
and the integrability condition (5) of twistor spinors with $\displaystyle{\not}D\psi_a=0$, we obtain
\begin{eqnarray}
\displaystyle{\not}dF&=&\bigg[\frac{1}{2}(K_c.u,e^c.\psi_a)-\frac{1}{2}(K_c.u,e_a.\psi^c)\nonumber\\
&&+\frac{\cal{R}}{4}(u,\psi_a)-(u,\nabla_{X_c}\nabla_{X_a}\psi^c)\bigg]e^a\nonumber\\
&=&\bigg[\frac{1}{2}(e^c.K_c.u,\psi_a)-\frac{1}{2}(e_a.K_c.u,\psi^c)\nonumber\\
&&+\frac{\cal{R}}{4}(u,\psi_a)-(u,\nabla_{X_c}\nabla_{X_a}\psi^c)\bigg]e^a.
\end{eqnarray}
We can use the equality $(e_a.K_c.u,\psi^c)=(u,K_c.e_a.\psi^c)$ and calculate the term $e^c.K_c$ as
\begin{eqnarray}
e^c.K_c&=&\frac{1}{n-2}\left(\frac{\cal{R}}{2(n-1)}e^c.e_c-e^c.P_c\right)\nonumber\\
&=&-\frac{\cal{R}}{2(n-1)}\nonumber
\end{eqnarray}
where we have used $e^c.e_c=n$ and $e^c.P_c={\cal{R}}$. Finally, the quantity $\displaystyle{\not}dF$ is found as follows:
\begin{eqnarray}
\displaystyle{\not}dF=\left[(u,\frac{n-3}{4(n-1)}{\cal{R}}\psi_a-\frac{1}{2}K_c.e_a.\psi^c-\nabla_{X_c}\nabla_{X_a}\psi^c)\right]e^a.\nonumber\\
\end{eqnarray}
This means that, to obtain a Maxwell field defined as in (39), $\psi_a$ of the Rarita-Schwinger field has to satisfy the following condition:
\begin{eqnarray}
\nabla_{X_c}\nabla_{X_a}\psi^c=-\frac{1}{2}K_c.e_a.\psi^c+\frac{n-3}{4(n-1)}{\cal{R}}\psi_a.
\end{eqnarray}
On the other hand, from the definition of the curvature operator, we have $\nabla_{X_c}\nabla_{X_a}\psi^c=\nabla_{X_a}\nabla_{X_c}\psi^c+R(X_c,X_a)\psi^c$. By using the action of the curvature operator on a spinor as $R(X_c,X_a)\psi^c=\frac{1}{2}R_{ca}.\psi^c$ and the property (23), one can write the condition as follows:
\begin{equation}
(R_{ba}+K_b.e_a).\psi^b=\frac{n-3}{2(n-1)}{\cal{R}}\psi_a
\end{equation}
which is automatically satisfied in a flat background. The constant coefficient on the right-hand side turns out to be $\frac{1}{6}$ in four dimensions.

\section{Symmetry operators}

Construction of spin raising and spin lowering operators gives way to write down symmetry operators for massless Rarita-Schwinger fields. A symmetry operator is an operator that acts on a solution of an equation and gives another solution of it. By starting with a Rarita-Schwinger field $\Psi$ and applying spin lowering and spin raising operators one after the other, one can find another Rarita-Schwinger field $\Psi'$ via a twistor spinor $u$. So, one can construct the symmetry operators of massless Rarita-Schwinger fields in terms of twistor spinors. However, this construction subjects to some constraints which arise in the procedures of spin raising and lowering.

Let us consider a massless Rarita-Schwinger field $\Psi=\psi_a\otimes e^a$ in four dimensions which satisfies (21) and (22) with an extra condition
\begin{equation}
(R_{ba}+K_b.e_a).\psi^b=\frac{1}{6}{\cal{R}}\psi_a.
\end{equation}
By using a twistor spinor $u$ that satisfies (4) with an extra condition
\begin{equation}
d\left[(u,\psi_a)e^a\right].u=0,
\end{equation}
we can construct the symmetry operators in the following way. Since $\Psi$ satisfies (49), we have a well-defined spin lowering procedure from spin-$\frac{3}{2}$ to spin-1 as in Sec. 3.B and can construct a source-free Maxwell field as follows:
\begin{eqnarray}
F&=&d\left[(u,\psi_a)e^a\right]\nonumber\\
&=&\left[\frac{1}{n}(e_a.\displaystyle{\not}Du,\psi_b)+(u,\nabla_{X_a}\psi_b)\right]e^a\wedge e^b.
\end{eqnarray}
We can use this source-free Maxwell field for spin raising to another spin-$\frac{3}{2}$ massless Rarita-Schwinger field. So, we can write the new spin-$\frac{3}{2}$ field from $F$ in (51) by the procedure in Sec. 3.A as
\begin{eqnarray}
\Psi'&=&\left(\nabla_{X_a}F.u-\frac{1}{n}F.e_a.\displaystyle{\not}Du\right)\otimes e^a\nonumber\\
&=&\bigg[\left\{\frac{1}{n}\nabla_{X_a}(e_b.\displaystyle{\not}Du,\psi_c)+\nabla_{X_a}(u,\nabla_{X_b}\psi_c)\right\}\nonumber\\
&&(e^b\wedge e^c).u\bigg]\otimes e^a\nonumber\\
&&-\bigg[\frac{1}{n}\left\{\frac{1}{n}(e_b.\displaystyle{\not}Du,\psi_c)+(u,\nabla_{X_b}\psi_c)\right\}\nonumber\\
&&(e^b\wedge e^c).e_a.\displaystyle{\not}Du\bigg]\otimes e^a
\end{eqnarray}
and by using the twistor equation (4) and the integrability condition (5), we obtain
\begin{eqnarray}
\Psi'&=&\bigg[\bigg\{\left(u,\nabla_{X_a}\nabla_{X_b}\psi_c+\frac{1}{2}K_a.e_b.\psi_c\right)\nonumber\\
&&+\frac{1}{n}\left(\displaystyle{\not}Du,e_a.\nabla_{X_b}\psi_c+e_b.\nabla_{X_a}\psi_c\right)\bigg\}(e^b\wedge e^c).u\bigg]\otimes e^a\nonumber\\
&&-\frac{1}{n}\bigg[\left\{(u,\nabla_{X_b}\psi_c)+\frac{1}{n}(\displaystyle{\not}Du,e_b.\psi_c)\right\}\nonumber\\
&&(e^b\wedge e^c).e_a.\displaystyle{\not}Du\bigg]\otimes e^a.
\end{eqnarray}
Since $u$ satisfies (50), $\Psi'$ is a massless Rarita-Schwinger field. So, we construct a symmetry operator between massless Rarita-Schwinger fields $\Psi\rightarrow L_u\Psi=\Psi'$ subject to some extra constraints. The symmetry operator $L_u$ constructed from a twistor spinor $u$ can be deduced from (53). It can also be deduced from (53) that the eigen-tensor spinors of the operator $L_u$ which are satisfying the condition $L_u\Psi=k\Psi$ for a constant $k$ and a Rarita-Schwinger field $\Psi=\psi_a\otimes e^a$ correspond to the solutions of the following equality:
\begin{eqnarray}
&&\bigg[\bigg\{\left(u,\nabla_{X_a}\nabla_{X_b}\psi_c+\frac{1}{2}K_a.e_b.\psi_c\right)\nonumber\\
&&+\frac{1}{n}\left(\displaystyle{\not}Du,e_a.\nabla_{X_b}\psi_c+e_b.\nabla_{X_a}\psi_c\right)\bigg\}(e^b\wedge e^c).u\bigg]\otimes e^a\nonumber\\
&&-\frac{1}{n}\bigg[\left\{(u,\nabla_{X_b}\psi_c)+\frac{1}{n}(\displaystyle{\not}Du,e_b.\psi_c)\right\}\nonumber\\
&&(e^b\wedge e^c).e_a.\displaystyle{\not}Du\bigg]=k\psi_a
\end{eqnarray}
which is not a trivial equation to solve.

\section{Conclusion}

We construct a solution generating technique for massless spin-$\frac{3}{2}$ Rarita-Schwinger fields by using source-free Maxwell fields and twistor spinors. A spin raising operator that maps a solution of the source-free Maxwell equations to a solution of the massless Rarita-Schwinger equation in terms of a twistor spinor is found for all even dimensional Ricci-flat backgrounds which is the requirement for the gauge invariance of the massless Rarita-Schwinger equation. A spin lowering operator that maps a solution of the massless Rarita-Schwinger equation to the solution of the source-free Maxwell field is also obtained in four dimensions with an extra constraint depending on the curvature characteristics of the background. From these spin raising and lowering procedures, a symmetry operator between massless Rarita-Schwinger fields is also constructed.

One can also investigate the construction of spin raising and lowering operators for spin-$\frac{3}{2}$ fields in more general spacetimes. For example, one can search the possibilities to construct spin changing operators via gauged twistor spinors which are generalizations of twistor spinors to Spin$^c$ spinors and can exist on more general backgrounds. This can extend the solution generating concept discussed in this paper to general cases. Moreover, the construction of spin raising and lowering operators for spin-2 and higher spin fields may also be investigated by similar procedures. Because of the consistency problems in the interactions of massless higher spin fields with nontrivial gravitational backgrounds, some restrictions may appear in the construction of spin raising and lowering operators for higher spin fields. These restrictions may reduce the backgrounds to the constant curvature spacetimes such as anti-de Sitter and flat backgrounds.

\begin{acknowledgments}
\"{O}. A. is supported by the Scientific and Technological Research Council of Turkey (T\"{U}B\.{I}TAK) Research Project No.118F086.
\end{acknowledgments}

\begin{appendix}
\section{}
Clifford algebra and spinor identities in physics literature are extensively written in terms of gamma matrices and abstract indices. Since the Clifford and exterior calculus notations are used in the papers that include previous calculations about the topic of the paper, we prefer to use this notation in our paper to have a direct connection with the previous results. The calculations are easier in this notation which is also more economic and elegant. However, it can easily be transformed into the language of gamma matrices and abstract indices. In this appendix, we give a summary of transformation rules between two notations and give the basic formulas in the paper in terms of gamma matrices.

In a flat Lorentzian spacetime, the gamma matrices satisfy the following Clifford algebra identity
\begin{equation}
\gamma^{\mu}\gamma^{\nu}+\gamma^{\nu}\gamma^{\mu}=2\eta^{\mu\nu}
\end{equation}
where $\eta^{\mu\nu}$ is the flat Lorentzian metric and $\mu,\nu$ are flat space indices. In a curved spacetime, the coframe basis 1-forms $e^a$ can be written in terms of coordinate components as $e^a=e^a_{\mu}dx^{\mu}$ where $e^a_{\mu}$ are called tetrad components and $e_a^{\mu}$ are the inverse tetrad. However, for the Clifford bundle, $e^a$ corresponds to the basis of the Clifford algebra and can be written in terms of gamma matrices as $e^a=e^a_{\mu}\gamma^{\mu}$. The curved space gamma matrices are defined in terms of tetrad components as follows:
\[
\gamma^a=e^a_{\mu}\gamma^{\mu}
\]
and they satisfy the following Clifford algebra identity:
\begin{equation}
\gamma^a\gamma^b+\gamma^b\gamma^a=2g^{ab}
\end{equation}
where $g^{ab}$ is the inverse metric and $a,b$ are curved space indices. So, the Clifford algebra basis $e^a$ and the curved space gamma matrices $\gamma^a$ are identical to each other.

The Dirac operator defined in (2) can be written in terms of gamma matrices as
\[
\displaystyle{\not}D=e^a.\nabla_{X_a}=e^a_{\mu}\gamma^{\mu}\nabla_a=\gamma^a\nabla_a
\]
where we have used $\nabla_{X_a}=\nabla_a$ in terms of abstract indices and omit the Clifford product notation. In this way, the twistor equation in (4) corresponds to
\begin{equation}
\nabla_au=\frac{1}{n}\gamma_a\displaystyle{\not}Du.
\end{equation}

A $p$-form $\alpha$ as an element of the Clifford bundle can be written in terms of the Clifford algebra basis as follows:
\[
\alpha=\frac{1}{p!}\alpha_{a_1a_2 ... a_p}\gamma^{a_1a_2 ... a_p}
\]
where $\gamma^{a_1a_2 ... a_p}$ corresponds to the antisymmetric combination of $\gamma^{a_1}\gamma^{a_2} ... \gamma^{a_p}$. For example, we have $\gamma^{ab}=\frac{1}{2}(\gamma^a\gamma^b-\gamma^b\gamma^a)$. So, the action of a $p$-form $\alpha$ on a spinor $\psi$ via the Clifford product $.$ is given in the following form:
\begin{equation}
\alpha.\psi=\frac{1}{p!}\alpha_{a_1a_2 ... a_p}\gamma^{a_1a_2 ... a_p}\psi.
\end{equation}
Then, the integrability conditions of the twistor equation given in (5) and (7) are written as
\begin{eqnarray}
\nabla_a\displaystyle{\not}Du&=&\frac{n}{2}K_{ab}\gamma^bu\\
C_{abcd}\gamma^{cd}u&=&0
\end{eqnarray}
while (6) remains unchanged. Here $K_{ab}$ are the components of the Schouten tensor and $C_{abcd}$ are the components of the conformal (Weyl) tensor.

For any two spinor fields $u$ and $v$ and a $p$-form $\alpha$, the spinor inner product $(\,,\,)$ has the following property given in the equation above (11):
\begin{eqnarray}
(u,\alpha_{a_1a_2 ... a_p}\gamma^{a_1a_2 ... a_p}v)=\nonumber\\
(-1)^{\lfloor p/2\rfloor}(\alpha_{a_1a_2 ... a_p}\gamma^{a_1a_2 ... a_p}u, v).\nonumber
\end{eqnarray}
So, for a massless spin-0 field $\phi$, we can obtain a massless spin-$\frac{1}{2}$ field via the spin raising operator given in (10)
\begin{equation}
\psi=(\partial_a\phi)\gamma^au+\frac{1}{2}\phi\displaystyle{\not}Du
\end{equation}
and from a massless spin-$\frac{1}{2}$ field $\psi$, we can obtain a massless spin-0 field via the spin lowering $\phi=(u,\psi)$. The symmetry operators given in (12) are
\begin{equation}
L_{u_1u_2}\psi=\partial_a(u_1,\psi)\gamma^au_2+\frac{1}{2}(u_1,\psi)\displaystyle{\not}Du_2.
\end{equation}
For the case of spin raising and lowering operators between massless spin-$\frac{1}{2}$ and spin-1 fields, we can write (13) and (14) as
\begin{eqnarray}
F&=&\gamma^au\otimes\overline{\nabla_a\psi}+\frac{1}{2}\displaystyle{\not}Du\otimes\overline{\psi}+\psi\otimes\overline{\displaystyle{\not}Du}\\
\psi&=&\frac{1}{p!}F_{a_1a_2 ... a_p}\gamma^{a_1a_2 ... a_p}u
\end{eqnarray}
and the symmetry operators in (15) corresponds to
\begin{eqnarray}
L_{u_1u_2}F&=&\frac{1}{p!}\gamma^au_2\otimes\overline{\nabla_a(F_{a_1a_2 ... a_p}\gamma^{a_1a_2 ... a_p}u_1)}\nonumber\\
&&+\frac{1}{2p!}\displaystyle{\not}Du_2\otimes\overline{F_{a_1a_2 ... a_p}\gamma^{a_1a_2 ... a_p}u_1}\nonumber\\
&&+\frac{1}{p!}F_{a_1a_2 ... a_p}\gamma^{a_1a_2 ... a_p}u_1\otimes\overline{\displaystyle{\not}Du_2}.\nonumber\\
\end{eqnarray}

A spin-$\frac{3}{2}$ field $\Psi=\psi_a\otimes\gamma^a$ is a massless Rarita-Schwinger field, if it satisfies the following equations:
\begin{eqnarray}
\mathbb{\displaystyle{\not}D}\Psi&=&0\\
\gamma^a\psi_a&=&0
\end{eqnarray}
where $\mathbb{\displaystyle{\not}D}=\gamma^a\nabla_a$ is the Rarita-Schwinger operator defined in (20). The Lorentz-type condition (23) corresponds to $\nabla^a\psi_a=0$. The spin raising operator from a spin-1 Maxwell field $F$ to obtain a massless spin-$\frac{3}{2}$ Rarita-Schwinger field via a twistor spinor $u$ given in (24) is written as
\begin{eqnarray}
\Psi&=&\bigg(\frac{1}{p!}\nabla_aF_{a_1a_2 ... a_p}\gamma^{a_1a_2 ... a_p}u\\
&&-\frac{1}{np!}F_{a_1a_2 ... a_p}\gamma^{a_1a_2 ... a_p}\gamma_a\displaystyle{\not}Du\bigg)\otimes\gamma^a.\nonumber
\end{eqnarray}
The condition (37) on the twistor spinor $u$ corresponds to
\begin{equation}
F_{a_1a_2 ... a_p}\gamma^{a_1a_2 ... a_p}u=0.
\end{equation}
The manipulations between (24) and (37) can be done in terms of gamma matrices in a similar manner to the calculations done in Sec. 3.A. For the spin lowering from spin-$\frac{3}{2}$ to spin-1, we have $A=(u,\Psi)$ given in (38) and the condition to obtain the Maxwell solution in (48) corresponds to
\begin{equation}
\left(\frac{1}{2}R_{bacd}\gamma^{cd}+K_{bc}\gamma^c\gamma_a\right)\psi^b=\frac{n-3}{2(n-1)}{\cal{R}}\psi_a
\end{equation}
where $R_{bacd}$ are the components of the Riemann tensor. The transformation rule between two massless Rarita-Schwinger fields given in (53) can be written as follows:
\begin{eqnarray}
\Psi' &=&\bigg[\bigg\{(u,\nabla_a\nabla_b\psi_c+\frac{1}{2}K_{ad}\gamma^d\gamma_b\psi_c)\nonumber\\
&&+\frac{1}{n}(\displaystyle{\not}Du,\gamma_a\nabla_b\psi_c+\gamma_b\nabla_a\psi_c+\gamma_b\nabla_a\psi_c)\bigg\}\gamma^{bc}u\bigg]\nonumber\\
&&-\frac{1}{n}\bigg[\bigg\{(u,\nabla_b\psi_c)+\frac{1}{n}(\displaystyle{\not}Du,\gamma_b\psi_c)\bigg\}\gamma^{bc}\gamma_a\displaystyle{\not}Du\bigg]\otimes\gamma^a.\nonumber\\
\end{eqnarray}

With the conventions defined in this appendix, all the derivations in the paper can be done by using gamma matrices in an equivalent manner to the Clifford and exterior calculus methods used in the paper.

\end{appendix}

%\references%

\end{document}